\documentclass[useAMS,usenatbib]{mn2e}

\usepackage{graphicx}
\usepackage{amsmath}
\title[Constraining broad-line regions]
{Constraining broad-line regions from time lags of broad emission lines
relative to radio emission}
\author[H. T. Liu, J. M. Bai, J. M. Wang and S. K. Li]{H. T. Liu$^{1,2}$\thanks{E-mail:
htliu@ynao.ac.cn; baijinming@ynao.ac.cn; wangjm@mail.ihep.ac.cn;
lisk@ynao.ac.cn}, J. M. Bai$^{1,2}$\footnotemark[1], J. M.
Wang$^{3,4}$\footnotemark[1]and
S. K. Li$^{1,2}$\footnotemark[1]\\
$^{1}$National Astronomical Observatories/Yunnan Observatory, Chinese Academy of Sciences, \\ Kunming, Yunnan 650011, China\\
$^{2}$Key Laboratory for the Structure and Evolution of Celestial
Objects, Chinese Academy of Sciences, \\ Kunming, Yunnan 650011, China\\
$^{3}$Key Laboratory for Particle Astrophysics, Institute of High
Energy Physics, Chinese Academy of Sciences,\\ 19B Yuquan Road, Beijing 100049, China\\
$^{4}$Theoretical Physics Center for Science Facilities, Chinese
Academy of Sciences, Beijing 100049, China}

\begin{document}

\date{Accepted . Received}

%\pagerange{\pageref{firstpage}--\pageref{lastpage}} \pubyear{2009}

\maketitle

\label{firstpage}

\begin{abstract}
In this paper, a new method is proposed to estimate the broad-line
region sizes of UV lines $R^{\rm{uv}}_{\rm{BLR}}$. It is applied
to 3C 273. First, we derive the time lags of radio emission
relative to broad emission lines Ly$\alpha$ and C IV by the ZDCF
method. The broad lines lag the 5, 8, 15, 22 and 37 GHz emission.
The measured lags $\tau^{\rm{uv}}_{\rm{ob}}$ are of the order of
years. For a given line, $\tau^{\rm{uv}}_{\rm{ob}}$ decreases as
the radio frequency increases. This trend results from the
radiative cooling of relativistic electrons. Both UV lines have a
lag of $\tau^{\rm{uv}}_{\rm{ob}}=-2.74^{+0.06}_{-0.25}$ yr
relative to the 37 GHz emission. These results are consistent with
those derived from the Balmer lines in Paper I. Second, we derive
the time lags of the lines Ly$\alpha$, CIV, H$\gamma$, H$\beta$
and H$\alpha$ relative to the 37 GHz emission by the FR/RSS Monte
Carlo method. The measured lags are
$\tau_{\rm{ob}}=-3.40^{+0.31}_{-0.05}$, $-3.40^{+0.41}_{-0.14}$,
$-2.06^{+0.36}_{-0.92}$, $-3.40^{+1.15}_{-0.20}$ and
$-3.56^{+0.35}_{-0.18}$ yr for the lines Ly$\alpha$, CIV,
H$\gamma$, H$\beta$ and H$\alpha$, respectively. These estimated
lags are consistent with those derived by the ZDCF method within
the uncertainties. Based on the new method, we derive
$R^{\rm{uv}}_{\rm{BLR}}=2.54^{+0.71}_{-0.35}$--$4.01^{+0.90}_{-1.16}$
and $2.54^{+0.80}_{-0.43}$--$4.01^{+0.98}_{-1.24}$ light-years for
the Ly$\alpha$ and CIV lines, respectively. Considering the
uncertainties, these estimated sizes are consistent with those
obtained in the classical reverberation mapping for the UV lines
and the Balmer lines. This indicates that their emitting regions
are not separated so large as in the classical mapping of the UV
and optical lines. These results seem to depart from the
stratified ionization structures obtained in the classical
mapping.

\end{abstract}

\section{INTRODUCTION}
Based on the photoionization assumption and the time lags between
both variations of broad emission lines and continuum,
reverberation mapping observations are able to determine the sizes
of broad-line regions (BLR) for type 1 active galactic nuclei
(AGNs) \citep[see e.g.]
[]{b30,b15,b48,b16,b14,b13,b32,b33,b34,b45}. The classical
reverberation mapping is very successful in estimates of the BLR
sizes. In these reverberation mapping observations, the stratified
ionization structures of the BLRs are found for various types of
broad emission lines, such as the Balmer lines and the HeII and
HeI lines seen in NGC 5548 \citep[e.g.][]{b7} and Mrk 110
\citep{b19,b18}, and the Balmer lines and the Ly$\alpha$ and C IV
lines seen in 3C 273 \citep{b29}. The higher ionized lines respond
systematically faster to the continuum variations. The separation
of emission regions for these lines spans a large range. The
higher ionized lines are emitted at the smaller distances from the
ionizing continuum source.

The photoionization assumption requires that the frequency of
ionizing continuum must not be lower than those of the emission
lines produced through the photoionization process
\cite[e.g.][]{b30}. The ultraviolet (UV) and optical continua are
used as the ionizing continuum in these reverberation mapping
observations. This treatment leads to underestimation of the BLR
sizes $R_{\rm{BLR}}$, as seen in e.g. 3C 273 \citep{b29}. For the
Balmer lines of 3C 273, the BLR sizes derived from the UV
continuum around 1300 $\rm{\AA}$ are between 2 and 4 times larger
than those from the optical continuum around 5000 $\rm{\AA}$.
Furthermore, their estimates are in excellent agreement with each
other \citep{b29}, while in \citet{b16} the H$\alpha$ lag is 60
per cent larger than that of H$\gamma$. It seems to be appropriate
to regard the UV continuum as the ionizing continuum of the Balmer
lines. For the UV lines Ly$\alpha$ and C IV, it should be
appropriate to regard the extreme-ultraviolet (EUV) and soft X-ray
continua as the ionizing continuum. \citet{b48} regarded the EUV
photons as the ionizing source of H$\beta$. \citet{b47} showed
that the soft X-rays are better to represent the ionizing flux.
The photons above 912 $\rm{\AA}$ are believed to be the main
sources of line formation via photoionization, and they should be
related to the emission lines seen in the near-ultraviolet (NUV)
and optical regimes \citep{b41}. Thus it seems to be inapposite of
using the UV continuum as the ionizing continuum of the UV lines
to estimate the BLR sizes.

Disturbances in the central engine are likely transported outward
along the relativistic jets. This was supported by observations
that dips in the X-ray emission are followed by ejections of
bright superluminal knots in the radio jets of AGNs
\citep{b23,b6,b4}. The events in the central engine will have a
direct effect on the events in the radio jets
\citep[e.g.][]{b23,b6}. According to the reverberation mapping
model \citep[e.g.][]{b5}, the events in the central engine also
have a direct effect on the events in the BLR through the
photoionization process. Thus it is expected that there should
exist correlations between both variations of the broad emission
lines from the BLR and the radio emission from the relativistic
jet aligned with the line of sight. Especially, the correlations
have time lags of the broad lines relative to the radio emission
and the time lags are related to the position of radio emitting
region $R_{\rm{radio}}$ \citep{b21}. Both optical and UV lines
should give the same $R_{\rm{radio}}$ from their time lags
relative to the radio emission. In this paper, we give a method to
estimate the BLR sizes of the UV lines from the BLR sizes of the
optical lines and their time lags relative to the radio emission.

\section{Method}
According to equation (7) in Paper I, we have a relation between
$R_{\rm{radio}}$, $R_{\rm{BLR}}$ and $\tau_{\rm{ob}}$
\begin{equation}
R_{\rm{radio}}=\frac{R_{\rm{BLR}}+\frac{c \langle
\tau_{\rm{ob}}\rangle}{1+z}}{\frac{c}{v_{\rm{d}}}-\cos \theta},
\end{equation}
where $c$ is the speed of light, $v_{\rm{d}}$ is the travelling
speed of disturbances down the jet, $\theta$ is the viewing angle
of the jet axis to the line of sight and $\langle
\tau_{\rm{ob}}\rangle \equiv \tau_{\rm{ob}}$ is the measured time
lag of the radio emission relative to the broad lines. Both the
optical and UV lines should give the same $R_{\rm{radio}}$. Hence,
we have $R^{\rm{uv}}_{\rm{BLR}}+c
\tau^{\rm{uv}}_{\rm{ob}}/(1+z)=R^{\rm{opt}}_{\rm{BLR}}+c
\tau^{\rm{opt}}_{\rm{ob}}/(1+z)$ that gives
\begin{equation}
R^{\rm{uv}}_{\rm{BLR}}=R^{\rm{opt}}_{\rm{BLR}}+\frac{c}{1+z}(\tau^{\rm{opt}}_{\rm{ob}}-\tau^{\rm{uv}}_{\rm{ob}}),
\end{equation}
where $R^{\rm{uv}}_{\rm{BLR}}$ is the BLR size of the UV lines, $R^{\rm{opt}}_{\rm{BLR}}$ is the
BLR size of the optical lines, $\tau^{\rm{uv}}_{\rm{ob}}$ is the measured time lag of the radio
emission relative to the UV lines and $\tau^{\rm{opt}}_{\rm{ob}}$ is the measured time lag of the
radio emission relative to the optical lines.

For the UV and optical broad emission lines, one can estimate the
time lags of lines by using the radio emission as the common
reference. As $\tau^{\rm{uv}}_{\rm{ob}}$ and
$\tau^{\rm{opt}}_{\rm{ob}}$ are measured, one can estimate
$R^{\rm{uv}}_{\rm{BLR}}$ from $R^{\rm{opt}}_{\rm{BLR}}$, which is
obtained in the reverberation mapping for the optical lines with
an appropriate ionizing continuum. This new method can improve the
classical reverberation mapping, especially on blazars that have
strongly beamed emission at optical, UV and X-ray bands and also
have strong radio emission.

\section{APPLICATION TO 3C 273}
The flat spectrum radio quasar 3C 273 was first identified as a
quasar at redshift $z=0.158$ by \citet{b36}. It is one of the best
studied AGNs in all bands \citep[e.g.][]{b46,b43,b39}. It is also
one of the bright extragalactic objects in the $\gamma$-ray sky.
Its jet is one-sided, with no signs of emission from the
counterjet side \citep{b44}. The blue-bump observed in 3C 273 is
attributed to thermal continuum emission from the inner accretion
disc \citep{b38}. Fe K$\alpha$ lines observed in 3C 273 are shown
to be from an accretion disc around a supermassive black hole
\citep{b50,b42}. The supermassive black hole accreting material to
form accretion disc powers the central engine in this object.

\subsection{Data and analysis of time lags}
This paper makes use of data from the 3C 273 database hosted by
the ISDC \footnote{http://isdc.unige.ch/3c273/} \citep{b43}. We
consider the radio light curves used in Paper I. Light curves of
broad UV lines Ly$\alpha$ and C IV are taken from \citet{b28}. The
sampling rates of the Ly$\alpha$ and C IV lines are around 6 times
per year, which are comparable with those of the Balmer lines used
in Paper I. First, we employ the same analysis method for the UV
lines as the Balmer lines H$\alpha$, H$\beta$ and H$\gamma$ used
in Paper I.

As in Paper I, the z-transformed discrete correlation function
\citep[ZDCF;][]{b1} is used to analyze the time lags that are
preferred to be characterized by the centroid $\tau_{\rm{cent}}$
of the ZDCF. The centroid time lag $\tau_{\rm{cent}}$ is computed
by using all the points with correlation coefficients $r \geqslant
0.8 r_{\rm{max}}$, where $r_{\rm{max}}$ is the maximum of
correlation coefficients in the ZDCF bumps closer to the zero-lag.
The calculated ZDCFs between the radio and broad-line light curves
are presented in Fig. 1. The horizontal and vertical error bars in
Fig. 1 represent the 68.3 per cent confidence intervals in the
time lags and the relevant correlation coefficients, respectively.
The ZDCFs in Fig. 1 have a common significant feature, i.e. the
negative lag closer to the zero-lag. All the ZDCF bumps closer to
the zero-lag have a good profile. The measured time lags are
listed in Table 1. The centroid $\tau_{\rm{cent}}$ is calculated
by $\tau_{\rm{cent}}=\sum\tau(i)r(i)/\sum r(i)$, where $\tau(i)$
and $r(i)$ are the values of the $i$th data pair with $r \geqslant
0.8 r_{\rm{max}}$. The errors of $\tau_{\rm{cent}}$ are calculated
by $\Delta \tau^{\pm}_{\rm{cent}}=\{\sum[\Delta \tau^{\pm}(i)
r(i)+\tau(i)\Delta r^{\pm}(i)]\sum r(i)-\sum \tau(i)r(i) \sum
\Delta r^{\pm}(i)\}/[\sum r(i)]^2$, where $\Delta \tau^{\pm}(i)$
and $\Delta r^{\pm}(i)$ are the relevant errors of $\tau(i)$ and
$r(i)$, respectively.

Our results show that the UV line variations lag the radio
variations of 5, 8, 15, 22 and 37 GHz (see Fig. 1). The measured
time lags are of the order of years (see Table 1). For a given
line, the relevant time lags generally decrease as radio frequency
increases from 5 to 37 GHz. This trend most likely results from
the radiative cooling of relativistic electrons (see Paper I).
These results are consistent with those obtained in Paper I.
Hereafter, $\tau_{\rm{cent}}$ is equivalent to $\tau_{\rm{ob}}$.

The uncertainties of each point in the ZDCFs only take into
account the uncertainties from the measurements by Monte Carlo
simulation. Thus the uncertainties in the cross-correlation
results will be estimated by using the model-independent FR/RSS
Monte Carlo method described by \citet{b31}. The quantities we
concerned are related with the time lags of Ly$\alpha$, CIV,
H$\alpha$, H$\beta$ and H$\gamma$ relative to the 37 GHz emission.
Thus we only recompute the uncertainties of time lags for these
five lines relative to the 37 GHz emission. The median of
distribution of the time lags estimated by Monte Carlo simulations
of 1000 runs is taken to be the time lag of line relative to the
37 GHz emission. The confidence intervals are estimated on the
basis of the distribution of the time lags simulated. The
calculated results are listed in Table 2. The time lags are
presented with centroid and peak values. For these five lines,
their time lags are well consistent with each other within $\pm
1\sigma$. These time lags of H$\alpha$, H$\beta$ and H$\gamma$
listed in Table 2 are well consistent with those listed in Table 1
of Paper I. The good agreements between these time lags estimated
by the FR/RSS method and the ZDCF method confirm the reliability
of the results of Paper I. Here, the uncertainties of time lags
are larger than those estimated in the ZDCF method. This indicates
underestimation of the uncertainties in the ZDCF method. Thus the
uncertainties presented in Table 2 will be used in the relevant
calculations of uncertainties.

\begin{figure}
\centering
\includegraphics[angle=0,scale=.4]{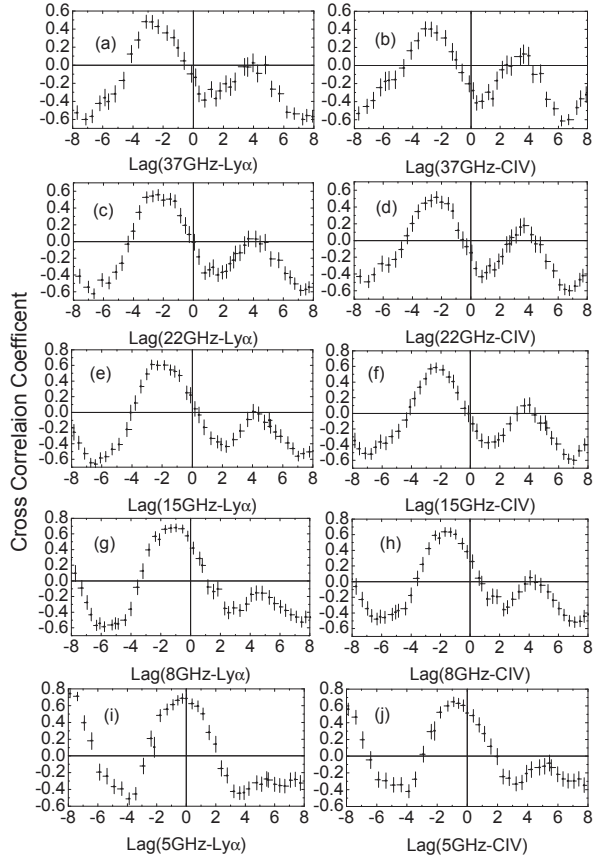}
 \caption{ZDCFs between Ly$\alpha$ and $(a)$ 37, $(c)$ 22, $(e)$ 15, $(g)$ 8, and $(i)$ 5 GHz;
 ZDCFs between C IV and $(b)$ 37, $(d)$ 22, $(f)$ 15, $(h)$ 8 and $(j)$ 5 GHz. The x-axis is in units of yr.
 }
  \label{fig1}
\end{figure}

\begin{table*}
\centering
 \begin{minipage}{100mm}
  \caption{Time lags between emission lines and
radio emission. The sign of time lag is defined as
$\tau_{\rm{cent}}= t_{\rm{radio}}-t_{\rm{line}}$. Time lags are in
units of yr.}
  \begin{tabular}{lrrrrr}

   \hline

 Lines& 5 GHz & 8 GHz&15 GHz &22 GHz&37 GHz\\
\hline

C IV&$-0.93^{+0.06}_{-0.21}$&$-1.67^{+0.04}_{-0.18}$&$-2.29^{+0.05}_{-0.22}$&$-2.34^{+0.05}_{-0.23}$&$-2.74^{+0.06}_{-0.25}$\\
Ly$\alpha$&$-0.24^{+0.05}_{-0.20}$&$-1.16^{+0.05}_{-0.22}$&$-2.08^{+0.04}_{-0.20}$&$-2.17^{+0.06}_{-0.23}$&$-2.74^{+0.06}_{-0.25}$\\

\hline

\end{tabular}
\end{minipage}
\end{table*}

\begin{table}
\centering
 \begin{minipage}{85mm}
  \caption{Time lags between emission lines and 37 GHz emission. The sign of time lag is defined as $\tau=
t_{\rm{radio}}-t_{\rm{line}}$. Time lags are in units of yr. The
numbers in brackets give the $\pm 1\sigma$, $\pm 2\sigma$ and $\pm
3\sigma$ confidence intervals.}
  \begin{tabular}{lrr}

   \hline

Lines & $\tau_{\rm{cent}}$ & $\tau_{\rm{peak}}$  \\
\hline

Ly$\alpha$&$-3.40\/\ \{^{+0.31}_{-0.05}\/\ \{^{+0.68}_{-0.19}\/\ \{^{+1.12}_{-1.00}$&$-3.40\/\ \{^{+0.10}_{-0.10}\/\ \{^{+0.90}_{-0.10}\/\ \{^{+2.00}_{-1.10}$\\

CIV &$-3.40\/\ \{^{+0.41}_{-0.14}\/\ \{^{+0.95}_{-0.51}\/\ \{^{+1.42}_{-1.05}$&$-3.40\/\ \{^{+0.20}_{-0.10}\/\ \{^{+1.00}_{-1.00}\/\ \{^{+1.90}_{-1.10}$\\

H$\gamma$&$-2.06\/\ \{^{+0.36}_{-0.92}\/\ \{^{+0.51}_{-1.54}\/\ \{^{+0.56}_{-2.68}$&$-1.90\/\ \{^{+0.30}_{-1.50}\/\ \{^{+0.40}_{-1.80}\/\ \{^{+0.40}_{-3.40}$\\

H$\beta$&$-3.40\/\ \{^{+1.15}_{-0.20}\/\ \{^{+1.90}_{-0.44}\/\ \{^{+2.06}_{-0.90}$&$-3.40\/\ \{^{+1.80}_{-0.30}\/\ \{^{+1.90}_{-0.50}\/\ \{^{+2.00}_{-1.10}$\\

H$\alpha$&$-3.56\/\ \{^{+0.35}_{-0.18}\/\ \{^{+0.86}_{-0.95}\/\ \{^{+0.96}_{-1.14}$&$-3.60\/\ \{^{+0.20}_{-0.10}\/\ \{^{+1.00}_{-1.00}\/\ \{^{+1.00}_{-1.60}$\\

\hline

\end{tabular}
\end{minipage}
\end{table}

\subsection{Estimation of BLR sizes}
Both UV lines, Ly$\alpha$ and C IV, have a time lag of
$\tau^{\rm{uv}}_{\rm{ob}}=-2.74$ yr relative to the 37 GHz
emission (see Table 1). Based on the typical size of optical lines
H$\alpha$, H$\beta$ and H$\gamma$ $R^{\rm{opt}}_{\rm{BLR}}=2.70$
light-years (ly) \citep{b29}, $\tau^{\rm{opt}}_{\rm{ob}}=-2.86$ yr
obtained in Paper I for the Balmer lines,
$\tau^{\rm{uv}}_{\rm{ob}}=-2.74$ yr and $z=0.158$, we obtain
$R^{\rm{uv}}_{\rm{BLR}}=2.60$ ly from equation (2). Without
considering the uncertainties, $R^{\rm{uv}}_{\rm{BLR}}=2.60$ ly
seems to be larger than those values of 1.2--1.9 ly obtained in
the reverberation mapping \citep{b29}. Furthermore, the lines
Ly$\alpha$ and C IV have the same time lag of
$\tau^{\rm{uv}}_{\rm{ob}}=-2.74$ yr relative to the 37 GHz
emission. These BLR sizes estimated by equation (2) for the lines
Ly$\alpha$ and C IV are in excellent agreement with each other.
These above results are based on the ZDCF method.

From equation (2), we have the uncertainty transfer expression
$\Delta R^{\rm{uv}}_{\rm{BLR}}=\Delta
R^{\rm{opt}}_{\rm{BLR}}+(\Delta \tau^{\rm{opt}}_{\rm{ob}}+ \Delta
\tau^{\rm{uv}}_{\rm{ob}})c/(1+z)$, which is the maximum
uncertainty transfer expression and will maximize $|\Delta
R^{\rm{uv}}_{\rm{BLR}}|$ due to possible combinations of $\Delta
R^{\rm{opt}}_{\rm{BLR}}$, $\Delta \tau^{\rm{opt}}_{\rm{ob}}$ and
$\Delta \tau^{\rm{uv}}_{\rm{ob}}$. Based on $\tau_{\rm{cent}}$
listed in Table 2, equation (2) and $R^{\rm{opt}}_{\rm{BLR}}$ of
\citet{b29}, we estimate $R^{\rm{uv}}_{\rm{BLR}}$ for the lines
Ly$\alpha$ and CIV in virtue of the lines H$\alpha$, H$\beta$ and
H$\gamma$. The transferred uncertainties are estimated by the
uncertainty transfer expression with the $\pm 1\sigma$, $\pm
2\sigma$ and $\pm 3\sigma$ confidence intervals of
$\tau_{\rm{cent}}$ and $R^{\rm{opt}}_{\rm{BLR}}$. The estimated
results are presented in Table 3. These estimated
$R^{\rm{uv}}_{\rm{BLR}}$ are consistent with each other within
$\pm 1 \sigma$. There seems to be a "flaw" in the $3\sigma$
uncertainties of $R^{\rm{uv}}_{\rm{BLR}}$ in the cases of the
lines H$\beta$ and H$\gamma$, for that the $3\sigma$ uncertainty
makes $R^{\rm{uv}}_{\rm{BLR}}$ to be negative (see Table 3). In
fact, this "flaw" just means that within $3\sigma$ uncertainty
$R^{\rm{uv}}_{\rm{BLR}}$ is consistent with zero. These above
results are based on the model-independent FR/RSS Monte Carlo
method \citep{b31}, which contains the DCF method \citep{b8}.

These $R^{\rm{uv}}_{\rm{BLR}}$ estimated from the ZDCF method are
consistent with those from the FR/RSS method within the
uncertainties. The BLR sizes of the lines Ly$\alpha$ and CIV
derived from the classical mapping are consistent with those
presented in Table 3 within about $\pm 2\sigma$ \citep[see Table 1
of][]{b29}. These agreements indicate for both UV lines that these
estimated $R^{\rm{uv}}_{\rm{BLR}}$ are reliable and reasonable.
Our results indicate that the separation between the BLRs of the
lines Ly$\alpha$ and C IV are not so large as in the classical
mapping \citep[see Table 1 of][]{b29}. Our results also indicate
for 3C 273 that the BLRs of UV lines are blended with or very
close to those of optical lines \citep[see Table 4 of][]{b29}.
\begin{table*}
\centering
 \begin{minipage}{120mm}
  \caption{$R^{\rm{uv}}_{\rm{BLR}}$ estimated for the Ly$\alpha$ and C IV lines in the cases of
  the H$\alpha$, H$\beta$ and H$\gamma$ lines. $R^{\rm{uv}}_{\rm{BLR}}$ is in units of ly. The
numbers in brackets give the transferred uncertainties from the
$\pm 1\sigma$, $\pm 2\sigma$ and $\pm 3\sigma$ confidence
intervals of $R^{\rm{opt}}_{\rm{BLR}}$,
$\tau^{\rm{opt}}_{\rm{ob}}$ and $\tau^{\rm{uv}}_{\rm{ob}}$.}
  \begin{tabular}{lrrr}

   \hline

Lines & H$\alpha$ & H$\beta$ & H$\gamma$  \\
\hline

Ly$\alpha$ & $2.54\/\ \{^{+0.71}_{-0.35}\/\ \{^{+1.66}_{-1.34}\/\
\{^{+2.38}_{-2.41}$ & $2.58\/\ \{^{+1.45}_{-0.41}\/\ \{^{+2.64}_{-0.87}\/\ \{^{+3.46}_{-2.84}$ & $4.01 \/\ \{^{+0.90}_{-1.16}\/\ \{^{+1.74}_{-2.18}\/\ \{^{+2.45}_{-5.81}$ \\

CIV & $2.54\/\ \{^{+0.80}_{-0.43}\/\ \{^{+1.89}_{-1.62}\/\
\{^{+2.64}_{-2.45}$ & $2.58\/\ \{^{+1.54}_{-0.48}\/\ \{^{+2.87}_{-1.15}\/\ \{^{+3.72}_{-2.88}$ & $4.01\/\ \{^{+0.98}_{-1.24}\/\ \{^{+1.97}_{-2.46}\/\ \{^{+2.71}_{-5.85}$ \\

\hline

\end{tabular}
\end{minipage}
\end{table*}

\section{Discussion}
The soft X-rays and the EUV continuum might be suitable to be used
as the ionizing continuum for the Ly$\alpha$ and C IV lines. The
soft X-rays are better to represent the ionizing flux
\citep{b20,b47,b10}. \citet{b48} regarded the EUV photons as the
ionizing source of H$\beta$. The photons above 912 $\rm{\AA}$ are
believed to be the main sources of line formation via
photoionization and should be related to the emission lines seen
in the NUV and optical regimes \citep{b41}. However, the soft
X-ray data are not available for many AGNs \citep[e.g.][]{b48}.
For blazars, the central ionizing continua will be strongly
contaminated by the beamed emission from the relativistic jets, so
that the blue-bump is not observable for most of blazars.
\citet{b27} showed for 3C 273 that the optical continuum is
strongly contaminated by the beamed emission from the relativistic
jet and it appears unsuitable for studying the time lags between
the ionizing continuum and the lines. \citet{b29} argued that the
UV continuum is much closer to the ionizing continuum than the
optical continuum used by \citet{b16}. They used the 1300
$\rm{\AA}$ continuum to infer the time lags of the Balmer lines
and found better estimate for the optical BLR than \citet{b16}.
The underlying physical condition in the photoionization model is
that the frequencies of ionizing photons must not be lower than
those of the emission lines generated via the photoionization
process. Thus it is appropriate to regard the 1300 $\rm{\AA}$
continuum as the ionizing continuum of the H$\alpha$, H$\beta$ and
H$\gamma$ lines, because that its frequency is higher than those
of the lines. However, it should not be appropriate to regard the
1300 $\rm{\AA}$ continuum as the ionizing continuum of the
Ly$\alpha$ line, because that its frequency is less than that of
the line.

There are several fundamental assumptions made in the
reverberation mapping, and one of them is that there is a simple,
although not necessarily linear, relationship between the
observable continuum and the ionizing continuum that is driving
the lines \citep{b30}. The assumption cannot be tested directly.
The close correspondence between continuum and line light curves
gives us some confidence that this assumption is valid at some
level of approximation \citep{b30}. However, observations indicate
a complicated relationship between variations of optical/UV and
EUV/X-ray bands, e.g. Seyfert type 1 AGN NGC 4051 \citep{b37,b2}.
The complex X-ray, EUV, UV and optical correlations are explained
as a possible combination of X-ray reprocessing, thermal Compton
up-scattering of optical/UV seed photons and disturbances
propagating from outer (optically emitting) parts of accretion
disc into its inner (X-ray emitting) region \citep{b37,b2}. The
complex relationships between these bands indicate that it is
difficult to choose the ionizing continuum and its appropriate
agency. This new method we proposed to estimate the BLR size of
the UV lines could avoid the choice of the ionizing continuum and
its agency.

The complex relationships between the UV, EUV and X-ray continuum
light curves may depend on the origin of energies emitted at these
bands. If the UV continuum is produced via thermal reprocessing of
the X-rays, the X-rays will lead the UV photons. If the UV, EUV
and X-rays are generated via viscous dissipation in accretion
disc, the UV will lead the EUV and X-rays. The EUV and X-ray
emission might be from the same origin. Observations of NGC 5548
on $\sim$ ten days show no time lag of X-rays from EUV flux on
time scales longer than a day \citep{b12}. If the X-rays are
produced by the up-scattering of UV seed photons, e.g. the X-rays
are generated in hotter corona and the UV seed photons are from
disc, the UV variations will lead the X-rays. The lag of the
X-rays relative to the UV variations corresponds to the
light-travel time between the seed photon source and a Compton
up-scattering region. This case seems to be supported by
observations of NGC 3516 that the optical variations lead the
X-rays by $\sim$ 100 days \citep{b22}. Their monitoring lasted for
$\sim$ 550 days. They showed that the correlation signal at 100
days is entirely due to the slow (variability time scale $\ga$ 30
days) components of the light curves. During the whole monitoring
period, the more rapidly changing components of the light curves
are uncorrelated at any lag. The light-travel distance of this lag
of X-rays relative to optical variations is much larger than the
BLR size of $\sim$ 11 light-days \citep{b49}. These outbursts,
which dominate this lag of $\sim$ 100 days, in the optical and
X-ray light curves have a variability time scale of the order of
$\sim$ 100 days. Observations of NGC 5548 on $\sim$ ten days show
that the time lag between the EUV and X-ray flux is negligible
relative to the light-travel time of the BLR size
\citep[see][]{b12}. Observations of NGC 3783 spanning 2 yr show
that the optical light curves lag the X-rays by 3--9 days
\citep{b3}. This delay points at optical variability produced by
X-ray reprocessing. This time lag is comparable to that of optical
lines relative to the continuum, $\sim$ 8 days \citep{b35,b40}.
The large time lags between these continuum bands should exist in
the long-term correlations, e.g. NGC 3516. The small lags should
exist in the short-term correlations, e.g. NGC 4151 and NGC 7469
\citep{b9,b25}. The light curves used to estimate the lags for NGC
4151 and NGC 7469 span time intervals $\la$ 30 days. It is
possible that the light curves used to measure the lags do not
have enough time spans, so that the large lags cannot be measured
even if indeed they exist. These observational facts indicate
complex relationships between these continuum bands and between
the time lags within these bands and the BLR sizes. Thus the
choice of ionizing continuum would significantly influence the
time lags of emission lines relative to the observable continua,
e.g. 3C 273 \citep[see][]{b29}.

According to the results of this paper, if the UV BLR size is
taken to be 2.6 ly for 3C 273, which means that if we had known
the light curve of the ionizing continuum (say the X-ray light
curve), the time lag would have been 2.6 yr. On the other hand,
the UV lines are lagging the UV continuum by about 1.5 yr
\citep{b29}. This means that the UV continuum light curve must lag
the X-ray light curve by about 1.1 yr, which means that the UV
emitting place is far from the X-ray emitting place by 1.1 ly. The
UV continuum light curve has a characteristic variability time
scale $\sim$ 2.0 yr, which is estimated from the zero-crossing
time of the autocorrelation function (ACF) of its light curve. The
zero-crossing time of the ACF of light curve is a well-defined
quantity and is used as a characteristic variability time scale
\citep[e.g.][]{b26,b1,b11}. There is a simple relation between the
ACF and the first-order structure function that is used in
variability studies to estimate the variability time scale
\citep{b11}. If this characteristic time scale $\sim$ 2.0 yr
corresponds to the size of the UV emitting region in accretion
disc, the UV emitting position will be $\sim$ 1.0 ly far from the
place that emits the X-rays. This distance of $\sim$ 1.0 ly is
consistent with that value of $\sim$ 1.1 ly deduced above. Thus it
is a reasonable size for the accretion disc, wherein the place
that emits the UV continuum is far from the place that emits the
X-rays by 1.1 ly.

For the Balmer lines H$\alpha$, H$\beta$ and H$\gamma$ of 3C 273,
the BLR sizes derived from the 1300 $\rm{\AA}$ continuum are
between 2 and 4 times larger than those from the optical continuum
around 5000 $\rm{\AA}$ \citep{b29}. Furthermore, their estimates
are in excellent agreement with each other, while in \citet{b16}
the H$\alpha$ lag is 60 per cent larger than that of H$\gamma$.
Thus the lines Ly$\alpha$ and C IV should have larger BLR sizes as
the soft X-rays rather than the 1300 $\rm{\AA}$ continuum are
taken to be the ionizing continuum. However, the soft X-rays do
not have enough data for 3C 273 \citep{b28}. Our new method
resolve the problem to choose the ionizing continuum. Based on
equation (2), $R^{\rm{opt}}_{\rm{BLR}}=2.70$ ly,
$\tau^{\rm{opt}}_{\rm{ob}}=-2.86$ yr,
$\tau^{\rm{uv}}_{\rm{ob}}=-2.74$ yr and $z=0.158$, we obtain
$R^{\rm{uv}}_{\rm{BLR}}=2.60$ ly for the lines Ly$\alpha$ and C
IV. If the time lags of the lines derived by the FR/RSS method are
used to estimate the UV BLR sizes by virtue of the Balmer lines,
$R^{\rm{uv}}_{\rm{BLR}}=2.54$--4.01 ly. These estimated sizes are
consistent with each other within $\pm 1\sigma$. The BLR sizes of
the lines Ly$\alpha$ and CIV obtained in the classical mapping are
consistent with these estimated sizes within about $\pm 2 \sigma$
\citep[see Table 3 of this paper and Table 1 of][]{b29}.
Considering the uncertainties, the BLR sizes of the lines
H$\alpha$, H$\beta$ and H$\gamma$ are well consistent with these
estimated sizes of the lines Ly$\alpha$ and CIV \citep[see Table 4
of][]{b29}. These results indicate that the BLRs of UV lines are
blended with or very close to those of optical lines. Our results
seem to depart from the stratified ionization structures obtained
in the classical mapping.

The new method proposed in this paper is based on the method
proposed in Paper I. The key of the two methods requires that the
central disturbance signals are transported to the BLR and the
jets. This key requirement of the two methods is supported by
observational researches \citep{b23,b6,b4} and theoretical
researches \citep[e.g.][]{b5,b24,b17}. The central disturbance
signals transported to the BLR and the jets are likely the
physical link between both variations of the emission lines from
the BLR and the beamed emission from the relativistic jet aligned
with the line of sight.

In Paper I, we found for the broad Balmer lines that the lags for
a given line generally decrease as radio frequency increases and
that this trend results from the radiative cooling of relativistic
electrons. For the UV lines Ly$\alpha$ and CIV, there are the same
cases as in the Balmer lines (see Table 1). The measured lags
$\tau^{\rm{ob}}_{\rm{lag}}$ between the 5, 8, 15, 22 and 37 GHz
emission were compared with the differences of $\Delta
\tau_{\rm{cent}}$ between $\tau_{\rm{cent}}$ for the Balmer lines
(see Fig. 7 in Paper I). In this paper, we compare
$\tau^{\rm{ob}}_{\rm{lag}}$ with $\Delta \tau_{\rm{cent}}$ for the
UV lines in Fig. 2. The line of $\Delta
\tau_{\rm{cent}}=\tau^{\rm{ob}}_{\rm{lag}}$ is consistent with the
measured data points for both the Balmer lines and the UV lines
(see Fig. 2). This agreement further confirms that the trend, i.e.
the lags for a given line generally decrease as radio frequency
increases, results from the radiative cooling of relativistic
electrons.
\begin{figure}
\centering
\includegraphics[angle=-90,scale=.3]{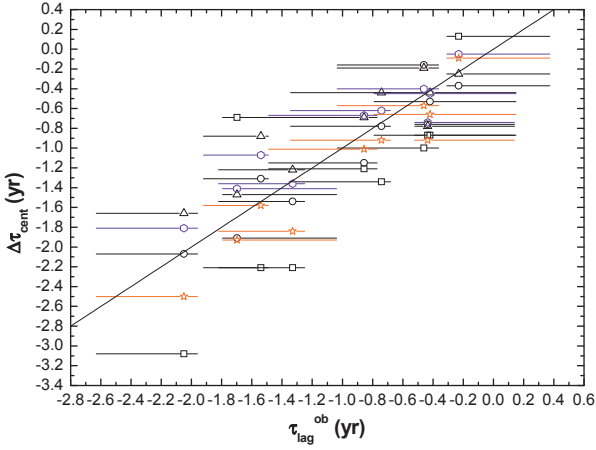}
 \caption{$\Delta \tau_{\rm{cent}}$ between $\tau_{\rm{cent}}$ for the five emission lines vs $\tau^{\rm{ob}}_{\rm{lag}}$.
 Circles present H$\alpha$, squares H$\beta$, triangles H$\gamma$, blue hexagons C IV and red pentacles Ly$\alpha$. Solid line is
 $\Delta \tau_{\rm{cent}}=\tau^{\rm{ob}}_{\rm{lag}}$.}
  \label{fig2}
\end{figure}

The broad-line light curves do not experience relativistic
shortening of variation time scales. The light curves generated by
a relativistic jet closely aligned to the line of sight experience
the relativistic shortening of variation time scales. The
relativistic effects would not have significant influence on the
estimates of time lags between the broad lines and the radio
emission (see Paper I). For testing the correctness of the time
lags measured by these ZDCFs, we compare the 37 GHz light curve
with the broad-line light curves moved horizontally and vertically
(see Fig. 3). For the negative lags used (see Fig. 1), the line
light curves are moved left by 2.7 yr. These moved line light
curves basically follow the variation trend of the radio light
curve (see Fig. 3). This indicates that the measured lags of
$\tau^{\rm{uv}}_{\rm{ob}}=-2.74$ yr are reliable and reasonable.
For the optical lines H$\alpha$, H$\beta$ and H$\gamma$, there are
the negative lags and the positive lags relative to the 37 GHz
emission (see Paper I). The lags should be positive or negative,
however the current data do not allow to discriminate between the
two cases for the Balmer lines. On the contrary, there are only
the negative lags for the Ly$\alpha$ and C IV lines relative to
the 37 GHz emission, because that these UV light curves span much
longer time intervals than do these optical line light curves. The
longer interval light curves do indeed resolve the problem of the
negative or positive lags emerging in Paper I, due to the shortage
of interval spans of these optical line light curves.
\begin{figure}
\centering
\includegraphics[angle=-90,scale=.3]{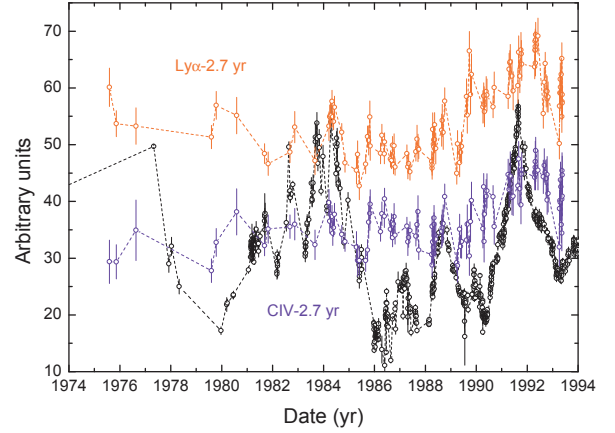}
 \caption{Comparison of the 37 GHz light curve with the Ly$\alpha$ and C IV line light curves
 moved along the x-axis and y-axis.}
  \label{fig3}
\end{figure}

\section{Conclusions}
In this paper, we propose a new method to estimate the BLR sizes
of the UV lines $R^{\rm{uv}}_{\rm{BLR}}$. It is applied to 3C 273.
We derive the time lags of the radio emission relative to the
broad emission lines Ly$\alpha$ and C IV. These broad lines lag
the 5, 8, 15, 22 and 37 GHz emission. The measured lags
$\tau^{\rm{uv}}_{\rm{ob}}$ are of the order of years. For a given
line, $\tau^{\rm{uv}}_{\rm{ob}}$ generally decreases as radio
frequency increases. This trend results from the radiative cooling
of relativistic electrons. Both the Ly$\alpha$ and C IV lines have
a time lag of $\tau^{\rm{uv}}_{\rm{ob}}=-2.74$ yr relative to the
37 GHz emission. These results are consistent with those derived
from the Balmer lines in Paper I. Based on
$\tau^{\rm{uv}}_{\rm{ob}}=-2.74$ yr,
$\tau^{\rm{opt}}_{\rm{ob}}=-2.86$ yr,
$R^{\rm{opt}}_{\rm{BLR}}=2.70$ ly and $z=0.158$, we obtain
$R^{\rm{uv}}_{\rm{BLR}}=2.60$ ly from equation (2). The time lags
estimated by the FR/RSS method are
$\tau^{\rm{uv}}_{\rm{ob}}=-3.40^{+0.31}_{-0.05}$ and
$-3.40^{+0.41}_{-0.14}$ yr for the lines Ly$\alpha$ and CIV,
respectively. These estimated lags are consistent with those lags
estimated by the ZDCF method. The time lags estimated by the
FR/RSS method are
$\tau^{\rm{opt}}_{\rm{ob}}=-3.56^{+0.35}_{-0.18}$,
$-3.40^{+1.15}_{-0.20}$ and $-2.06^{+0.36}_{-0.92}$ yr for the
lines H$\alpha$, H$\beta$ and H$\gamma$ relative to the 37 GHz
emission, respectively. These estimated lags are well consistent
with those lags estimated by the ZDCF method in Paper I. These
lags estimated by the FR/RSS method for the UV lines and the
Balmer lines are well consistent with each other within the
uncertainties. Based on these estimated lags,
$R^{\rm{opt}}_{\rm{BLR}}$ obtained in the classical mapping and
equation (2), we derive the BLR sizes of the UV lines. These
estimated sizes of Ly$\alpha$ are
$R^{\rm{uv}}_{\rm{BLR}}=2.54^{+0.71}_{-0.35}$,
$2.58^{+1.45}_{-0.41}$ and $4.01^{+0.90}_{-1.16}$ ly in the cases
of H$\alpha$, H$\beta$ and H$\gamma$, respectively. These
estimated sizes of CIV are
$R^{\rm{uv}}_{\rm{BLR}}=2.54^{+0.80}_{-0.43}$,
$2.58^{+1.54}_{-0.48}$ and $4.01^{+0.98}_{-1.24}$ ly in the cases
of H$\alpha$, H$\beta$ and H$\gamma$, respectively. These
estimated sizes are well consistent with each other within $\pm
1\sigma$. These estimated sizes of the UV lines are consistent
with those UV BLR sizes obtained in the classical mapping within
about $\pm 2 \sigma$ \citep[see Table 3 of this paper and Table 1
of][]{b29}. Considering the uncertainties, these estimated sizes
of the UV lines are well consistent with the BLR sizes of the
Balmer lines obtained in the classical mapping \citep[see Table 4
of][]{b29}. These results indicate that the BLRs of UV lines are
blended with or very close to those of optical lines. Our results
seem to depart from the stratified ionization structures obtained
in the classical mapping.

\section*{Acknowledgments}

We are grateful to the anonymous referee for constructive comments
and suggestions leading to significant improvement of this paper.
H.T.L. thanks the West PhD project of the Training Programme for
the Talents of West Light Foundation of the CAS, and the National
Natural Science Foundation of China (NSFC; Grant 10903025) for
financial support. J.M.B. acknowledges the support of the NSFC
(Grants 10973034 and 10778702). J.M.W. is supported by the NSFC
(Grant 10733010). J.M.B. and J.M.W. acknowledge the support of the
973 Program (Grant 2009CB824800).

\label{lastpage}

\end{document}